\documentstyle[12pt]{article}
\newcommand{\vp}{\varphi}

\begin{document}
\begin{center}
{\bfseries  Notes on relativistic superfluidity   and gauge/string
duality
}
\vskip 5mm
H. Verschelde$^{\dag}$, V. I. Zakharov$^\ddag$
\vskip 5mm
{\small {\it $^\dag$ Ghent University, Department of Physics and Astronomy\\
Krijgslaan, 281-S9, 9000 Gent, Belgium}} \\
{\small {\it $^\ddag$ITEP, B. Cheremushkinskaya 25, Moscow, 117218 Russia,\\
Max-Planck Institut f\"ur Physik, 80805 M\"unchen, Germany}}\\
\end{center}
\vskip 5mm
\centerline{\bf Abstract}
{We consider selected topics of relativistic superfluidity within gauge/string duality. Non-relativistically, the only conservation law relevant  to the hydrodynamic approximation is the energy-momentum  conservation.     Relativistically, one has to introduce   an extra conserved $U(1)$ current and an extra three-dimensional scalar field  which is condensed.  Finding out a proper $U(1)$ symmetry     becomes a crucial point. We emphasize that in dual models   there     do arise extra $U(1)$ symmetries associated with wrapping of the  strings around (extra) compact directions in Euclidean space-time.   Moreover, if the geometry associated  with an extra compact dimension is cigar-like then the corresponding  $U(1)$ symmetry could well be spontaneously broken.   The emerging Goldstone particle survives in the hydrodynamic limit. A specific point is that the chemical potential conjugated    with the corresponding $U(1)$ charge is vanishing. Within the standard approach the vanishing chemical  potential implies no superfluidity.  We argue that an exotic liquid, introduced recently in the literature, with vanishing energy density and non-vanishing pressure represents a viable description of the liquid associated with 3d Goldstone particles in Euclidean space-time.  Since it lives on the stretched membrane, it describes energy-momentum transport in the deep infrared.  We discuss briefly possible applications to the quark-gluon plasma. }
\newpage
\section{Introduction}
The discovery of the quark-gluon plasma  at RHIC \footnote{For details, discussions and references  see, e.g. reviews \cite{teaney}.}  made a profound impact on theoretical developments.  From a theoretical point of view, the most important observation  is that the plasma appears to be a  relativistic quantum liquid. This conclusion follows  primarily  from the  low numerical value \footnote{ For a recent analysis of the data see \cite{romatschke}.} of the ratio of the shear viscosity $\eta$   to the entropy density $s$,  close to the conjectured   lowest bound  \cite{son1}
\begin{equation}\label{low}
  \frac{\eta}{s}\ge\frac{1}{4\pi}~,
\end{equation}
which is a kind of uncertainty principle for hydrodynamics.

This discovery boosted applications of holographic models to condensed-matter systems, and further unification of methods of elementary-particles and condensed-matter physics, for review see \cite{herzog1}. There exist not many models of quantum liquids,  and superfluidity is a natural first candidate.  And indeed, the hydrodynamics of relativistic superfluidity has been elaborated recently in much more detail than during  preceding decades.  Recent analyses  of superfluidity within holographic approaches can be found, in particular, in \cite{minwala}.

There are various theoretical signatures of superfluidity. For our purposes, the most relevant one is the appearance of a pole in the static correlator of densities of momentum. In more detail, consider the correlator
\begin{equation}\label{0i}
G_R^{0j,0i}(k)~\equiv~i\int d^4xe^{-ikx}\theta(t)\langle|T^{0j}(x),T^{0i}(0)|\rangle ~,
\end{equation}
where $T^{0i}~(i=1,2,3)$ are components of the energy momentum tensor. Furthermore, concentrate on the case of zero frequency, $k_0=0$, and space momentum tending to zero, ${\bf k}\to 0$. Then there is a pole contribution to the $G_R^{0j,0i}(k)$ (for a recent relativistic derivation and references see \cite{yarom}):
\begin{equation}\label{poleterm}
\lim_{k_i\to 0}{G_R^{0j,0i}(k_i, k_0=0)}~=~{k_ik_j\over k^2}\rho_s\mu~~,
\end{equation}
where $\rho_s$ is the density of the superfluid component and $\mu$ is the chemical potential.

Note an important difference in description of superfluidity in the  relativistic and non-relativistic cases. In the non-relativistic case the introduction of the chemical potential $\mu$ does not imply existence of a conserved Noether current.  Indeed, non-relativistically the number of particles is conserved simply because  there are no annihilation graphs. As a result, the chemical potential $\mu$ can be introduced for neutral particles. Relativistically, one does introduce a Noether current $J_{\mu}$ and assumes  \cite{superfluidity} that there exists a 3d complex field $\Phi$ condensed in the vacuum,
\begin{equation}\label{vev}
\label{u1}<\Phi>~\neq~0~~.
\end{equation}
Then the phase  $\phi$ of the field $\Phi$ becomes a 3d Goldstone field manifested in  Eq (\ref{poleterm}).  

One of our central points is that stringy models of QCD or Yang-Mills theories do have extra $U(1)$ symmetries which are not obvious at all in terms of the field theoretic formulation of these theories \cite{verschelde}. Such symmetries apply then in the infrared limit of the Yang-Mills theories. These effective $U(1)$ symmetries arise if there exist  compact coordinates, and the corresponding quantum number is the winding number of  a string around  a compact dimension, see, e.g., \cite{polchinskibook}. The existence of at least one such compact dimension is guaranteed at non-zero temperature. We have in mind   the Euclidean time $\tau$. As is well known, $\tau\sim\tau+1/T$. The corresponding winding number is a topological quantum number in stringy models. Near the temperature $1/\beta_H$ of the Hagedorn phase transition the stringy modes with the winding number $n_{w}=\pm 1$ become massless \cite{kogan,atick}. At temperatures $1/\beta$ close to $1/\beta_H$ these modes represent a complex scalar field $\Phi_{therm}$, called thermal scalar,  with the mass
\begin{equation}\label{tachyon}
m^2_{therm}~\approx~{\beta_H(\beta_H-\beta)\over2\pi^2(\alpha^{'})^2}~.
\end{equation} Moreover, at temperatures above $1/\beta_H$ the thermal scalar could well condense
\begin{equation}
<\Phi_{therm}>~\neq~0,
\end{equation}
as speculated in a number of papers, see in particular \cite{atick,polchinski,rabinovici,silverstein}. Then the phase $\phi_{therm}$ of the field $\Phi_{therm}$ would become a 3d massless field, apparently inducing superfluidity \cite{verschelde}. Namely, the static correlator (\ref{poleterm}) can be calculated directly in the Euclidean space-time and exhibits a pole  term \cite{verschelde}.

Here we come to a problem, however. Indeed, it is very natural, if not imperative, to assume that the Yang-Mills plasma is neutral with respect to the quantum number associated with the winding around the compact Euclidean time $\tau$. In other words,
\begin{equation}
\mu_{\tau-winding}~=~0.
\end{equation}
Then, according to (\ref{poleterm}) the coefficient in front of the pole term should vanish. On the other hand, direct calculation of the static correlator in the Euclidean time $\tau$ gives a non-vanishing result since $$|\partial_{\tau}\phi_{therm}|~=~2\pi T~,$$ where $T$ is the temperature.

Thus, one is encouraged to look for   alternative models  in the Minkowskian time which would avoid proportionality of the coefficient in front of the pole term to $\mu$, see (\ref{poleterm}). Remarkably enough, such an alternative seems to be provided by the exotic liquid introduced in Ref. \cite{skenderis}. In the equilibrium the energy density for such a liquid is vanishing while pressure is non-vanishing. In other words, the components of the energy-momentum tensor in the equilibrium are given by:
\begin{equation}\label{liquid}
<T_{00}>~=~0,~~<T_{0i}>=0,~~<T_{ij}>~=~p\delta_{ij}~.
\end{equation}
Moreover, as we will argue, the static correlator introduced above does have a pole term:
\begin{equation}\label{poleterm1}
\lim_{k_i\to 0}{\int d\tau \exp{(-ik_ix_i)}<T^{0i}(x,\tau), T^{0j}(0)>}~\sim~(const)
{k_ik_j\over k^2}~~,
\end{equation}
with a nonvanishing  $(const)$.
 
The liquid considered   possesses features similar to a superfluid.
First , the correlator (\ref{poleterm1}) does exhibit a pole, as in the superfluid case.
   Also, the ratio $\eta/s$ for this liquid is calculated holographically  in \cite{skenderis} and   equals the conjectured lowest bound $\eta/s=1/4\pi$. 
    
    The price for avoiding proportionality of the pole term to the chemical potential $\mu$ (compare (\ref{poleterm})) is that the effective lagrangian introduced in \cite{skenderis} is highly non-linear. It is worth emphasizing that the introduction of the liquid (\ref{liquid}) in Ref. \cite{skenderis} is not at all motivated by considering superfluidity. Instead, the authors pursue duality between (or, better to say,  mapping of) the solutions of the Einstein equation in $(d+1)$ dimensions and solutions of non-relativistic Navier-Stokes equations in $d$ dimensions, for more details see \cite{keeler}.

The outline of the paper is as follows. In Sect. 2 we review the standard, two-component picture of superfluidity, emphasizing the points which we are going to modify later. In Sect. 3 we discuss light scalars which arise within holographic models of Yang-Mills theories. In Sect. 4 we argue that the liquid introduced in Ref. \cite{skenderis} might describe a superfluid component in case of vanishing chemical potential.

\section{Standard two-component hydrodynamics}

\subsection{Set of equations}

A superfluid is usually defined as a liquid in which a Bose condensate is formed. Therefore, the superfluidity is commonly described   within a two-component picture with two densities, $\rho_n, \rho_s$ corresponding to the normal liquid and the condensate. The basic new feature, compared to the  ordinary liquid is existence of a new light degree of freedom. In the original non-relativistic set up this degree of freedom is the slowly varying   phase of the wave function of $N$ identical particles  (with momentum {\bf p}=0):
\begin{equation}\label{psi}
\Psi({\bf r,}t)~=~\sqrt{N}\exp(-i\phi({\bf r},t))~.
\end{equation}
The time dependence of the phase $\phi$ is fixed as $\partial_t\phi=\mu$ (Josephson equation, valid in the system at rest with respect to the normal component). The Josephson equation follows from the (non-relativistic) canonical commutation relation
\begin{equation}
[\rho_s({\bf x}),\phi({\bf y})] = i\delta({\bf x- y})
\end{equation}
and the fact that the Hamiltonian of the system depends on $\mu$ through the common relation $H=H_0-\mu N$. In particular, the pure state
\begin{equation}\label{purestate}
\Psi({\bf r,}t)~=~\sqrt{N}\exp(-i{\bf P}\cdot {\bf r})\exp(i\mu t)
\end{equation}
describes a superfluid component flowing with momentum ${\bf  P}$ with respect to the normal component (which is at rest).

To rewrite hydrodynamics in a relativistically invariant way, one introduces the four-vector velocity of an element of the liquid (its normal component) $u^{\mu}$ so that, for example, the Josephson  equation becomes $u^{\mu}\partial_{\mu}\phi~=~\mu$. From the perspective of the present paper, the most important change is that relativistically the number of particles is not conserved any longer and the chemical potential now enters the Hamiltonian through $H=H_0-\mu Q$ where $Q$ is a novel, conserved charge associated with a Noether current $J^{\nu}$.

A fully relativistic version of superfluid hydrodynamics was elaborated first in Refs \cite{lebedev,carter}.  Later it was noticed \cite{superfluidity} that contribution of the $\phi$ field to the $U(1)$ conserved current $J^{\nu}$ and the energy-momentum tensor $T^{\nu\sigma}$ can be kept the same  as for ordinary massless (3d Goldstone) field. In this form, the contributions of the condensate component to the  energy-momentum tensor $T^{\nu\sigma}$ and the current $J^{\nu}$  look similar to the standard field theoretic expressions.  Adding the contribution of the normal component  and neglecting first dissipation one gets \cite{yarom}:
\begin{eqnarray}\label{son}
T^{\nu\sigma}~=~(\epsilon+P)u^{\nu}u^{\sigma}~+P~\eta^{\nu\sigma}+
{\rho_s\over \mu}\partial^{\nu}\phi\partial^{\sigma}\phi\\ \nonumber
J^{\nu}~=~\rho_nu^{\nu}+{\rho_s\over \mu}\partial^{\nu}\phi\\ \nonumber
u^{\mu}\partial_{\mu}\phi+\mu~=~0\\ \nonumber
dP~=~sdT+\rho_nd\mu-{\rho_s\over 2 \mu} (\partial_{\mu}\phi)^2 ~~,
\end{eqnarray}
where we included also a thermodynamic relation which contains   a new  term proportional to  $ {1\over 2}(\partial_{\mu}\phi)^2 $.  This is a reflection of the    independence of the 4-velocities of the normal and  superfluid components. Furthermore, $\rho_s, \rho_n$ are the condensate and normal densities, respectively, $T$ is the temperature,  $P$ is the pressure,  $s$ is the entropy density,  $\mu$ is the chemical potential, $\epsilon$ is defined as the Legendre transform of the pressure with respect to the temperature and chemical potential, $\epsilon=-P+Ts+\rho_n\mu$. Eqs (\ref{son}) are borrowed from Ref \cite{yarom} where the reader can find further details. Generalizations to include dissipation can be found in Refs. \cite{son1,minwala}. Note also that there exist various forms of the basic equations of the relativistic two-component hydrodynamics, see, e.g., \cite{son1}. To get convinced that a relativistic scheme describes superfluidity one evaluates the spectrum of excitations, like sound waves, second sound, see \cite{yarom} and references therein. The equations (\ref{son}) quoted above do lead to the spectrum characteristic for superfluidity. Another check is that entropy flow is associated only   with the normal component. In the ideal-liquid approximation (\ref{son}) one gets indeed
\begin{equation}\label{entropy}
\partial_{\mu}(su^{\mu})~=~0~.
\end{equation}
There is no entropy associated with the condensate component, as it should be since the condensate component corresponds to a pure quantum state (\ref{purestate}).

\subsection{Comments on the ratio $\eta/s$}

Note that for the condensate component both the viscosity $\eta$ and entropy density $s$ vanish so that the ratio $\eta/s$ is not defined. In the holographic models of the superfluidity the bound (\ref{low}) refers in fact to the normal component, for a recent analysis see   \cite{minwala}. If one sticks to the description of the superfluid component exclusively in terms of the pure state (\ref{purestate}) then the ratio $\eta/s$ remains undefined.  Physics-wise, one can appreciate the situation   in the following way. Imagine that one is monitoring superfluidity by measuring properties of the liquid. Then any measuring procedure would introduce decoherence and, as a result,
 produce  a normal component, with a non-vanishing entropy.
 In this way one would resolve the uncertainty in the value of the ratio $\eta/s$ in case of superfluidity. The textbook-presentation of the superfluid component with $\eta,s\equiv 0$ corresponds to another limiting procedure: there are no measurements on the superfluid liquid and the superfluid current persists for any long  time. The use of a holographic description in terms of black holes corresponds  to the former definition.

We are not aware of any explicit demonstration of the relation between   
measuring procedure and induced $\eta/s$ ratio. However, the paper in Ref. \cite{stodolsky} provides a relevant example of similar phenomena. Very briefly, one considers a two-level system in terms of the density matrix, $$\rho~=~{1\over 2}\big(I+{\bf P}\cdot {\bf \sigma}\big)~,$$ where $\sigma$ are the Pauli matrices and ${\bf P}$ is the polarization vector. There are measurements performed on the two-level system. As a result, $$\dot{{\bf P}}~=~{\bf V}\times {\bf P}-D{\bf P}_{tr}~~ $$ where the vector ${\bf V}$ describes evolution of the system in absence of measurements, the label ``tr'' means transverse to the $z$ axis, an $D$ stands for damping or decoherence.  If $D\neq 0$ a pure initial state develops into a mixed state, with corresponding increase in   the entropy. On the other hand, $D$ is proportional to the variance of current fluctuations  induced in the  measuring device, $$D~\sim~[\bar{{j^2}}-\bar{j}^2]~~,$$ where the coefficient of proportionality is related to the parameters of the device. Fluctuations, in turn are related to dissipation through the fluctuation-dissipation theorem, and we refer the reader for all details to the original paper \cite{stodolsky}.

\subsection{Static pole}

 As is mentioned in the Introduction, in  case of superfluidity certain static correlators reveal the presence of a 3d Goldstone particle. Consider first the static correlator of space components of the current $J^i$ (see Eqs (\ref{son})). It involves two structure functions:
 \begin{equation}\label{currentcurrent}
G_R^{i,j}(k_0=0, {\bf k})~\equiv~{k^ik^j\over {\bf k}^2}G_R^{J||}({\bf k})+
\Big(\delta^{ij}-{k^ik^j\over {\bf k}^2}\Big)G_R^{J\perp}({\bf k})~~.
\end{equation}
The low energy theorems for  (\ref{currentcurrent}) are very similar to the famous case of the correlator of two electromagnetic currents in case of superconductivity:
\begin{equation}\label{readily}
\lim_{{\bf k}\to 0}G_R^{J\perp}({\bf k}) ~=~-\rho_s/\mu,~~~\lim_{{\bf k}\to 0}G_R^{J||}({\bf k}) ~=~0~,
\end{equation}
where $\rho_s$ is the density of the superfluid component, see above. One readily derives (\ref{readily}) by noting that in the static case the field $\phi\sqrt{\rho_s/\mu}$ is normalized in the standard way, and has a $1/{\bf k}^2$ propagator in the momentum space.

The massless 3d exchange is also manifested in the static correlator of densities of the total momentum, see (\ref{0i}). Moreover, this correlator is a kind of more fundamental since in the non-relativistic limit superfluidity can exist even if there is no independent conserved current $J^{\mu}$ so that $J^{\mu}$ is proportional to $T^{0\mu}$.   In more detail,
\begin{equation}
G_R^{0i0j}(k_0=0,{\bf k})~=~{k^ik^j\over {\bf k}^2}G_R^{T||}({\bf k})+
\Big(\delta^{ij}-{k^ik^j\over {\bf k}^2}\Big)G_R^{T\perp}({\bf k})~.
\end{equation}
The low energy theorems, with inclusion \cite{yarom} of relativistic corrections, now read:
\begin{equation}
\lim_{{\bf k}\to 0}{G_R^{T\perp}({\bf k})}~=~-\big(sT+\mu\rho_n\big)~,
\lim_{{\bf k}\to 0}{G_R^{T||}({\bf k}}~=~-\big(sT+\mu(\rho_n+\rho_s)\big)~~,
\end{equation}
which also can be rewritten as (\ref{poleterm}).

For our purposes, it is crucial that the pole term (\ref{poleterm}) vanishes in the limit of vanishing chemical potential $\mu=0$.

 The static long-range interaction  is manifested both in the non-relativistic and relativistic set up. Within the context of the present paper it is important, however, that the mechanism of generating a 3d Goldstone particle is very different in these two cases. Non-relativistically, the existence of the pure state (\ref{purestate}) is in no way related to a spontaneous   breaking of any symmetry. Instead, it is rather due to initial conditions which assume the existence of N identical bosons \footnote{ As is well known, in case of superfluidity matrix elements of the annihilation and production operators $a_0^{\pm}$ of the particles with {\bf p}=0  are   the same in the leading $N$ approximation, $<a_0^{+}>\approx <a_0^{-}>$.  Sometimes this is taken as a signal of a spontaneous symmetry breaking.   However, the similarity is only formal since in case of superfluidity the number   of particles is rigorously conserved and the   approximate equality of the matrix elements  of $a_0^{\pm}$   is possible because fixing the chemical potential fixes   the number of particles only on average.  The $1/N$ corrections   to this approximate equality are explicitly taken into account by the Bogolyubov transformation.   The point was recently emphasized in \cite{shirkov}.   }.

\subsection{Extra scalar field}

In the relativistic case, one does introduce a complex scalar field and assumes a non-vanishing vacuum expectation value (\ref{vev}). There is no need for such a field in the non-relativistic limit. True,  Eqs (\ref{son}) do have a smooth match to the non-relativistic case, with chemical potential $\mu$ becoming essentially the mass of a particle and equation $\partial_{\mu}J^{\mu}=0$ becoming identical to the equation $\partial_{\mu}T^{0\mu}=0$. However, the field contents of the non-relativistic and relativistic versions of the theories do not match each other.

Since this point is crucial in the context of our paper let us quote the recently found holographic model of superfluidity \cite{herzog2,minwala}. It starts with a five-dimensional space:
\begin{equation}
ds^2~=~{L^2\over u^2}\Big(-f(u)du^2+dx^2+dy^2+dz^2+{du^2\over f(u)}\Big)~,
\end{equation}
where $f(u)=1-(u/u_h)^4$ and our 4d world corresponds to the boundary $u\to 0$. The bulk action $S_{bulk}$ apart from gravity contains a complex scalar $\Psi$ and gauge field $F_{ab}$:
\begin{equation}
S_{bulk}=\int d^5x\sqrt{-g}\Big[{1\over 2\pi^2}\big(R+{12\over L^2}\big)-{1\over 4e^2}F^{ab}F_{ab}
-m^2|\Psi|^2-B|\Psi|^4 -(1+A|\Psi|^2)|\partial\Psi-iA\Psi|^2 \Big],
\end{equation}
where $A,B$ are constants. The chemical potential is related to the boundary value of the $A_t$ component of the vector potential and the scalar condensate is determined by the boundary value of $\Psi$. For a certain range of parameters (essentially, for $\mu/T>2$) there exists a 3d scalar-field condensate. Note that the Lorentz invariance is violated from the very beginning by choosing a finite temperature.

The introduction of an extra scalar field in   the bulk allows to construct holographic models  of superfluidity. To apply this construction to specific systems, say the quark-gluon plasma, it is crucial to elaborate on the nature of the extra $U(1)$ symmetry.

\section{Stringy U(1) symmetries}

 \subsection{Thermal scalar }

 As is discussed in the preceding section, a prerequisite to having relativistic superfluidity is spontaneous breaking of a $U(1)$ symmetry. At first sight, this condition severely limits applications of superfluidity to realistic cases, say, to QCD. However, things change drastically in stringy models. Then, if there exists a compact coordinate, the winding number becomes an Abelian charge (see, e.g., \cite{polchinskibook}). Moreover, in the Euclidean signature existence of such a compact coordinate is granted, since the Euclidean time $\tau$ is periodic at finite temperature, $\tau~\sim~\tau~+~1/T$.

The possibility of spontaneous breaking of the corresponding $U(1)$ symmetry has been discussed intensely in the literature,   see, e.g., \cite{atick,kruczenski,alvarezgaume} and references therein. Historically, implications of the thermal $U(1)$ symmetry were discussed first in critical, $d=26$ dimensions and the lightest stringy mode with non-trivial wrapping number is called the thermal scalar \cite{kogan,atick}. Thus, we first review briefly the issue of the thermal scalar. The original set up for the thermal scalar does not seem nowadays to be a realistic framework to describe Yang-Mills theories. However, as we will argue  later some basic features of the thermal scalar survive in more realistic frameworks. What is even more amusing, there are good reasons to at least discuss seriously the possibility of a spontaneous breaking of the corresponding $U(1)$ symmetry, see, e.g., \cite{atick,kruczenski,alvarezgaume} and references therein.

 In more detail, it was realized \cite{kogan,atick} that at the Hagedorn temperature a complex scalar field $\Phi_{threrm}$, the so called thermal scalar, becomes massless, see Eq. (\ref{tachyon}). The Euclidean-time dependence of this field is fixed as
 \begin{equation}\label{thermal}
\Phi_{therm}(\tau, {\bf x})~=~\exp (i2\pi\tau T)\phi_{therm}({\bf x})~,
\end{equation}
where $\phi_{therm}({\bf x})$ is a 3d complex scalar field. The free energy for the thermal scalar is given by
\begin{equation}\label{free}
F~=~~-\beta\ln Z~=~-\beta \int_0^{\infty}{dL\over L (l_sL)^d/2}\exp(-m^2_{therm}Ll_s)~,
\end{equation}
where $L$ is the length of a closed trajectory, $d$ is the number of non-compact, spatial dimensions, $l_s$ is the string scale. Moreover, the free energy (\ref{free}) coincides with the partition function for a single static string with tension $1/2\pi \alpha^{'}$ and winding around Euclidean time once.

Equation (\ref{free}) is nothing else but the random-walk representation of the free energy for a complex scalar field with a {\it fixed} size of the walking step, $l_s$. If we had a fundamental scalar, then a representation similar to (\ref{free}) would be valid for any length of the step (or ``lattice spacing"). This fixation of the length of the step is the first indication that we are dealing with an effective scalar degree of freedom. For strings related to QCD, $l_s\sim (\Lambda_{QCD})^{-1}$.  Thus, we are discussing in fact an infrared phenomenon, or  strong-coupling dynamics. It is in this regime that one might hope that  dual, or stringy models apply.  

So far we considered strings in flat space. More relevant to the modern dual models are extra dimensions with non-trivial geometry. The results  concerning the thermal scalar do generalize from the flat to curved spaces, for a concise  presentation see \cite{kruczenski}. In the background metric $ds^2=G_{\tau\tau}d\tau^2+ G_{ij}dx^idx^j,$ and in the presence of a static dilaton $\Phi$ the effective action is
\begin{equation}\label{effective}
S_{\phi}~=~\beta\int d^dx\sqrt{G}e^{-2\Phi}\Big(G^{ij}\partial_i\phi^{*}\partial_j\phi+
{G_{00}\beta^2-\beta_H^2\over 4\pi^2\alpha^{'2}}\phi^{*}\phi\Big)
\end{equation}
where $\phi$ is a complex field depending on $d$ spatial coordinates. The action (\ref{effective}) allows to evaluate various spatial correlators at large distances within full string theory at temperature close (and below) the Hagedorn temperature $T_H$. In other words, at $T/\sqrt{G_{00}}\approx T_H$ string contributions are dominated by a single, light degree of freedom.

It is worth emphasizing that there exist two dual interpretations of the thermal scalar. One way to visualize it is that (\ref{tachyon}) refers to the mass of the mode once wrapped around the compact, Euclidean time direction. Another way is to demonstrate that partition function of a long string reduces to a random-walk representation of  a partition function associated with a single light scalar degree of freedom. It is sometimes said, therefore, that the thermal scalar is unphysical. The observation that long-distance correlators in string theory are calculable as an exchange of the thermal scalars emphasizes that the thermal scalar can be treated in the infrared as an ordinary field.

\subsection{Yang-Mills theories}

It is worth emphasizing that dominance of the thermal scalar was demonstrated at $T<T_H$. At $T>T_H$ and within the original set up of papers in Ref. \cite{kogan} there are no massless scalars left. Moreover, we are  interested in the deconfinement phase transition occurring at $T=T_c$ and $T_H=T_c$ only at critical dimensions, $d=26$ for the bosonic strings, see \cite{kogan}. Thus, our actual theoretical framework is modern dual models for Yang-Mills theories (at large $N_c$). It turns out, however, that the thermal-scalar related phenomena are generic and might well persist at temperatures $T>T_c$.

To substantiate this point, let us remind the reader of the dual model \cite{witten} for large-$N_c$ Yang-Mills theories which seems most viable. The metric at zero temperature reads as
\begin{eqnarray}\label{zerotemperature}
ds^2~=~\big({u\over R}\big)^{3/2}(-dt^2+\delta_ijdx^idx^j+ f(u)dx_4^2)\\\nonumber
+\big({u\over R}\big)^{-3/2}\Big({du^2\over f(u)}+u^2d\Omega_4^2\Big)\\
\end{eqnarray}
where~$$f(u)=1-{u_{\Lambda}^3\over u^3}~,~x_4~\sim~x_4+\beta_4,~~\beta_4~=~\big({4\pi\over 3}\big)\big({R^3\over u_{\Lambda}}\big)~$$ and the four-dimensional YM coupling is given by $g_{YM}~=~2(2\pi)^2g_sl_s/\beta_4$ and the 't Hooft coupling is defined as $\lambda~ \equiv~g_{YM}^2N_c$.

The topology of the space considered is $R^{3,1}\times D\times S^4$. In particular, the physical size of the circle goes  to zero smoothly as $u$ approaches the horizon $u_{\Lambda}$ from below. The background gives confinement in the dual gauge theory, as can be seen by computing the quark-antiquark potential in terms of the action of a string with endpoints on the boundary at $u\to ~\infty$. Confinement is also manifest from the  non-contractibility of the Polyakov loop in the Euclidean finite temperature backgrounds at $T<T_c$.

At large temperatures and in the Euclidean  time $\tau$ the metric is given by
\begin{eqnarray}\label{temperature}
ds^2~=~\big({u\over R}\big)^{3/2}(f(u)d\tau^2+\delta_{ij}dx^idx^j+dx_4^2)\\ \nonumber
+\big({u\over R}\big)^{-3/2}\Big({du^2\over f(u)}+u^2d\Omega_4^2\Big)~~,
\end{eqnarray}
where $f(u)$ is the same as above with replacement of $u_{\Lambda}$ by temperature horizon $u_T$. Note that now (at large $T$) the cigar type geometry is in the coordinates $(u,\tau)$ while the radius of the compact $x_4$-coordinate does not depend on $u$.

\subsection{Light states}

The main lesson from the holographic models (\ref{zerotemperature}) and (\ref{temperature}) is that in both cases there are cigar-type geometries. This, in turn implies that there exist light states which, generally speaking, should be taken into account in the low-energy dynamics. Usually, contribution of the winding modes is disregarded in the infrared limit. Indeed, by construction the model (\ref{zerotemperature}) applies only at distances $r\gg\beta_4$ and, naively, one can neglect the winding modes. However, the cigar-shape geometry makes this argumentation doubtful. The equation (\ref{tachyon}) for the mass of the thermal scalar illustrates the point well. Thus, a holographic cigar-shape geometry might indicate light states. In this section, we will review briefly whether this expectation is supported  by phenomenology.

In this context, it is worth emphasizing that, apart from the mass of a winding mode, the tension associated with non-perturbative defects, D-branes, can also vanish at the horizon.  Generically, for the probability $W_{defect}$  to observe   lower-dimension defects, or D-branes one has:
\begin{equation}
W_{defect}~\sim~\exp(-S_{defect})~,~~~S_{defect}~\sim~N_c~~,
\end{equation}
and this probability vanishes in the $N_c\to \infty$ limit.  However, in the models considered  some radii of extra dimensions are vanishing, see (\ref{zerotemperature}), (\ref{temperature}). Then, in the classical approximation there are defects whose action is vanishing on the horizon: $$S_{defect}(u\sim u_h)~\sim~N_c\cdot 0~~.$$ Thus, generally speaking,  such defects should be added 'by hand' to the standard set of light states. In fact, at temperatures close and higher than  the  temperature of the deconfinement phase transition $T_c$  some defects of this type have been already discussed in the literature quite in detail, see in particular \cite{pervye,gorsky,gorsky2}.

  Instantons represent  the simplest and best known example of such a ``defect".   It is well known that at $T>T_c$ instantons are suppressed exponentially in the large $N_c$ limit. On the other hand, instantons are infrared unstable and not suppressed at zero temperature. The holographic models above do reproduce these facts in a simple geometric language, in terms of the cigar-type geometry  \cite{pervye}. It is less known that there are further vacuum defects, in particular percolating  D2 branes. Percolation is interpreted as vanishing of the corresponding tension \cite{gorsky, gorsky2}.  Moreover, the continuum-theory predictions  based on the geometries  (\ref{zerotemperature}) and (\ref{temperature}) fit well the lattice data. All in all, one can say that there exists a compelling evidence that  the non-perturbative physics becomes three-dimensional at $T\ge T_c$ \cite{nakamura}. This is a direct reflection of the cigar-type geometry in the $(\tau,u)$ coordinates, see (\ref{temperature}). At zero temperature a generic consequence of the geometry (\ref{zerotemperature}) is that defects with non-trivial $\theta$-dependence have action vanishing in the infrared.

In this paper, we concentrate on hypothetical light particles associated with strings wrapped around the compact directions present in metrics (\ref{zerotemperature}) and (\ref{temperature}). At $T>T_c$ it is the Euclidean time and at $T<T_c$ it is the $x_4$ coordinate. Unlike the case of magnetic D-branes, phenomenology does not either produce any decisive  evidence in favor of light winding modes or rules them out.

Two further remarks are in order. First, from the naive geometrical point of view states associated with any non-zero winding number around the time direction become massless at the horizon. We assume, however, that the cigar-shape geometry is just another language for a vanishing scalar mass. In other words, we have to discuss only a single light state, not a whole tower of such states. Similar views were presented earlier in  somewhat different language \cite{rabinovici}. Second, it was speculated in \cite{silverstein} that the effective mass of the thermal scalar depends on the value of $u$ and a negative mass squared (or condensation) at $u\sim u_T$ cannot be detected in ``our world" which corresponds to  the ultraviolet limit $u\to \infty$. To our mind, condensation at the horizon,  $u \sim u_H$ fixes generically the mass scale of the condensate in our world, say, $<\Phi>^2\sim \Lambda_{QCD}^2$, and does not necessarily imply vanishing of the condensate. This is a standard viewpoint in the confining models.  On the other hand, it is true that there is no firm theoretical control over the physics in the infrared and one cannot rule out more exotic possibilities \footnote{Moreover, there is an example \cite{chernodub2} from the lattice measurements when the condensate is of order $v^2\sim\Lambda_{QCD}^3a$ where $a$ is the lattice spacing and indeed vanishes in the continuum limit $a\to 0$.}.

Winding modes emerge in terms of strings. We are unaware of an explicit expression for such modes in terms of the dual gauge theory. There is another, somewhat similar defect, the Polyakov line $l({\bf x})$, which is, to the contrary, formulated in terms of the 4d gauge field and refers to the integral over Euclidean time:
\begin{equation}\label{polyakov}
 l({\bf x})~\equiv~{1\over N_c}Tr{\bf L({\bf x})}~;~~{\bf L({\bf x})}~\equiv~P\exp\Big(ig\int_0^{1/T}A_0({\bf x},\tau) d\tau\Big)~.
 \end{equation}
 It was speculated that this defect becomes light at $T>T_c$  already in the original paper by Polyakov  and nowadays there are many models that utilize this idea,  see, e.g. \cite{pisarski} and references therein.

The path of integration in (\ref{polyakov}) is wrapped around Euclidean time and $l({\bf x})$  can be considered as an effective 3d field. Thus, there are similarities  to the thermal scalar. Moreover, if we consider a Polyakov line as an isolated  defect then it clearly has a 3d translational zero mode which is a 3d massless  field $\varphi_{P}$. However, because of the trace in Eq (\ref{polyakov})   the field $\varphi_P({\bf x})$ is a real field, unlike the thermal scalar which is    complex.  Moreover, if one assumes that the Polyakov line becomes dynamical and is condensed  at $T>T_c$ then  $$<\varphi_P>~\neq~0~, $$  and there is no massless state left in the spectrum   (while condensation of a complex thermal scalar implies a massless Goldstone particle).  In fact, there exist models for Yang-Mills medium build along these lines,  see \cite{deforcrand} and references therein.

In conclusion of this section we would like to add a point which  escaped discussion altogether, as far as we know. At $T=0$ the cigar-shape geometry in the $(u,x_4)$ coordinates might well imply the existence of a 4d massless Goldstone particle. Indeed, strings wrapped once over the compact $x_4$ direction would correspond to a complex 4d field $\Phi_{\theta}(x_{\mu}),$ with the topological charge $Q_{top}=\pm 1$. Condensation of such a field $$<\Phi_{\theta}>~\neq~0~~,$$ would imply existence of a 4d Goldstone particle coupled to the topological charge. As usual, the Goldstone particle would couple to the states with non-vanishing $U(1)$ charges. In the approximation considered, the thermal winding string modes   are heavy and non dynamical. However, the Goldstone field could couple also to  external fields carrying topological charge:
\begin{equation}
L(\phi_{\theta})~=~1/2(\partial_{\mu}\phi_{\theta})^2~+~ f_{\theta}\partial_{\mu}\phi_{\theta}K_{\mu} +...
\end{equation}
where $K_{\mu}$ is the topological Chern-Simons current and $f_{\theta}$  is a constant, $f_{\theta}\sim \Lambda_{QCD}$. In other words the Goldstone particle would play the role of an axion. Detailed consideration of such a possibility goes beyond the scope of the present paper.

\subsection{Reservations}

The basic observation which we are exploiting is that the models   (\ref{zerotemperature}), (\ref{temperature}) possess extra $U(1)$  symmetries in the ultraviolet  in terms of the winding states. Because of the cigar-type geometry,  however, these states can be unwound in the infrared. Generically,  this implies spontaneous symmetry breaking of the corresponding $U(1)$ symmetry  and, then,  the emerging Goldstone particle is a new light degree of freedom  surviving in the infrared.

 Clearly, there are many reservations to this conclusion. First, the geometries  are derived in the $N_c\to \infty$ limit and for realistic $N_c$ the would be  "Goldstone particles" could be massive (although with rather small masses).  Second, the models (\ref{zerotemperature}), (\ref{temperature}) do not  actually describe Yang-Mills theories in the ultraviolet. Thus, the  Goldstone particle might also be an artefact of the model.  Third, it was argued \cite{silverstein} that condensation of  the thermal scalar in the infrared might not be manifested in any  kind of measurable effect in our world.    Thus, we proceed to phenomenological manifestations of the hypothetical   light states. In the future,  the phenomenology could   decide, whether such states    exist.

\section{Superfluidity with $\rho_s=0$ ?}

\subsection{Two signatures of superfluidity}

 It is crucial for the present paper that there exist  two different signatures of superfluidity. The first one is derivable from the set of equations (\ref{son}). Reiterating the basic points: there is a medium with density of a conserved charge $q(x)$ ($\int q(x)d^3x=Q$). The total charge is shared by two components so that in the limit of low relative velocity of both components $$q({\bf x})=\rho_n({\bf x})+\rho_s{(\bf x})~$$ where the relative distribution of the total charge carried by $\rho_n,\rho_s$ depends on the temperature.  However, it is only the normal component with 4-velocity $u^{\mu}$ which carries entropy, $\partial_{\mu}(su^{\mu})=0$, and this is superfluidity.

The other signature is the appearance of a pole in the static correlator (\ref{poleterm}). In the non-relativistic set up  the two signatures are  manifestations of the same mechanism, that is formation of a condensate, (see Eq. (\ref{purestate})). Also relativistically, within the standard framework outlined in Sect. 2, both signatures are manifestations of the same mechanism. Indeed, the same superfluid density $\rho_s$ which carries no entropy according to (\ref{entropy}) controls also the strength of the pole term according to (\ref{poleterm}),
\begin{equation}\label{residue}
r_{pole}~=~\mu\rho_s~,
\end{equation}
where $r_{pole}$ is the residue of the pole and $\mu$ is the chemical potential conjugated to the charge $Q$. Note that for a real scalar field $\phi$  $$T_{0i}~\sim~\partial_0\phi\partial_i\phi~~,$$  and to evaluate $r_{pole}$ we treat the time derivative, $\partial_0\phi=\mu$  as a c-number while $\partial_i\phi$ is treated as an operator.

We will argue now that the stringy models considered above might represent a challenge to this wisdom.

\subsection{The problem to be solved}

 As an example, consider again the $U(1)$ symmetry associated with  the winding around the Euclidean time direction. The wave function of the thermal scalar is given by (\ref{thermal}). Assume now that the 3d complex field $\phi_{therm}$ develops a non-vanishing vacuum expectation value, $$<\phi_{therm}> ~=~v ~\neq~0~~,$$ and evaluate the correlator of the momentum densities. The (Euclidean) time derivative is treated again as  non-fluctuating, $|\partial_{\tau}\Phi_{therm}|~=~2\pi T v$,  where $T$ is temperature and $v$ is the 3d vacuum expectation value. The spatial derivative of the thermal scalar is treated as an operator, $\partial_i\Phi_{therm}=iv\partial_i\phi_{therm}$, where $\phi_{thermal}$ is a 3d massless field.

As a result, there arises a pole term in the Euclidean  static correlator of the momentum densities  with the pole residue equal to:
\begin{equation}\label{poleterm3}
r_{pole}~=~-T(2\pi T)^2v^2
\end{equation}
The central point is that the value of the residue is not proportional to   any chemical potential.   On the other hand, the analytic continuation of a  static correlator from Euclidean to Minkowski space seems to be trivial  (apart from an overall minus sign). And we seemingly come to contradiction  with (\ref{residue}).

 Moreover, it seems most natural to consider the case of a vanishing   chemical potential associated with the winding $U(1)$ symmetry,  $\mu_{\tau-winding}=0$. If there is no chemical potential then   we expect vanishing density of charge anyhow.   Concentrating on the superfluid  component, we thus expect:
 \begin{equation}\label{vanishing}
 \rho_s~=~0~,
 \end{equation}
which again is in variance with the standard description.

To summarize, in case of spontaneous breaking of a Euclidean  $U(1)$ symmetry   there exists 3d Goldstone particle manifested as a  pole in the correlator of the $T_{0i}$ components  of the energy-momentum tensor. From this point of view,   we expect superfluidity. On the other hand,  the density of  the condensate, $\rho_s$ is expected to vanish. Based on  the latter observation we would conclude that there is  no superfluidity. As we will argue now, the resolution of the  paradox  is that in the relativistic cases the two signatures of the superfluidity can in fact be separated.

 \subsection{An exotic liquid}

  Recently a new kind of holographic fluid was introduced in Ref. \cite{skenderis}. As we will argue, this fluid  actually satisfies the conditions specified in the preceding subsection: namely, it exhibits the pole (\ref{poleterm3}) and has vanishing density.

First, let us quote  the basic results of Ref. \cite{skenderis} relevant to our problem. Consider the following non-linear Lagrangian for a real scalar field $\vp$ \footnote{The dimensionful const $T$ is put to unity in what follows.}:
\begin{equation}\label{action}
S~=~T\int d^4x\sqrt{-\gamma}\sqrt{-(\partial\vp)^2}~,
\end{equation}
where $\gamma$ is the determinant of the metric tensor $\gamma_{ab}~,a,b=0,1,2,3$ defined as
\begin{equation}\label{metric1}
\gamma_{ab}dx^adx^b~=~-r_cd\tau^2+dx_idx^i~~,
\end{equation}
where the speed of light, $\sqrt{r_c}$ is arbitrary. The action (\ref{action}) describes an ideal liquid with 4-velocity $$u_a=\partial_a\vp/\sqrt{X}~, X~\equiv~-(\partial\vp)^2~~,~u_au^a=-1~.$$  We will also need the stress tensor:
\begin{equation}\label{tab}
 T_{ab}~=~\sqrt{X}\gamma_{ab}+{1\over \sqrt{X}}\partial_a\vp\partial_b\vp~.
 \end{equation}
Finally, the equilibrium corresponds to the solution
\begin{equation}\label{continuation}
 \vp_{equilibrium}~=~\tau~~,
\end{equation}
which breaks Lorentz invariance and defines the rest frame.  Note that in this subsection we stick to  the notations of the Ref. \cite{skenderis}  where further details can be found.

 \subsection{Similarities to superfluidity }

 Let us now map the properties of the exotic liquid introduced in \cite{skenderis}  to the properties of the liquid we are looking for.

 Imagine that we would try to continue directly the solution (\ref{thermal})  from Euclidean to Minkowski space. The solution (\ref{thermal}) oscillates  in the Euclidean time and, naively, in Minkowski space would have either  growing or dissipating solution which cannot correspond to equilibrium.  Thus, instead of the naive continuation we would look for a solution  in the Minkowski space which would reproduce the static correlator, see Eqs.  (\ref{poleterm}),  (\ref{poleterm3}). As we emphasized a few times, the time derivative of  the scalar field entering the components $T_{0i}$    of the energy-momentum tensor is treated classically. Therefore we would  look for the solution $\vp\sim\tau$ in the Minkowski space as well.   Therefore the ansatz (\ref{continuation}) fits very well our scheme  (although introduced in Ref. \cite{skenderis} for absolutely different  reasons).

 Next, the use of the square root $\sqrt{-(\partial\vp)^2}$ in the  action (\ref{action}) might look exotic. But this is exactly what is needed   to ensure (\ref{vanishing}). Indeed, the equilibrium form of the energy-momentum  tensor (\ref{tab}) is diagonal:
 \begin{equation}
 \Big(T_{ab}\Big)_{equilibrium}~=~(0,~p,~p,~p)~~,
 \end{equation}
 where the pressure $p$ is given by $$p~=~{1\over \sqrt{r_c}}~,$$   and $r_c$ is introduced in the metric (\ref{metric1}). Moreover, the form of the action (\ref{action}) was fixed in \cite{skenderis} exactly for the same reason of getting the energy density vanishing. (Although the motivation to have $\epsilon_{equilibrium}=0$ in \cite{skenderis} is absolutely different from ours).

Moreover, as far as one can use the action (\ref{action}) to describe the medium,  i.e. in the ideal-liquid approximation, the conservation of the energy-momentum tensor is granted, $\partial_{\mu}T^{\mu\nu}=0$ without introduction of entropy. In other words, entropy $$s=0~,$$is  the same as for the superfluid component in the ideal-liquid approximation  of Eqs. (\ref{son}). And the reason for the vanishing entropy is the same  as in case (\ref{son}). Indeed, the contribution of the scalar field $\phi$ to the current $J_{\mu}$ and energy-momentum tensor $T_{\mu\nu}$ can be borrowed from the standard action of the scalar field and this is the reason, why the superfluid component does not carry entropy if $\partial_{\mu}\phi$ is treated (upon normalization) as its velocity.

As the next step, the correlator of the momentum densities can readily be evaluated using (\ref{tab}) and treating the $\partial_{\tau}\vp$ as a classical field and $\partial_i\vp$ as an operator:
\begin{equation}
\lim_{k_i\to 0}{G_R^{0j,0i}(k_i, k_0=0)}~=~ {k_ik_j\over k^2}\sqrt{r_c}
\end{equation}
Fitting this result to (\ref{poleterm3}) allows to relate the parameter $r_c$ (which is arbitrary in the framework of \cite{skenderis}) to our 3d vacuum expectation value $v$.

Finally, it is shown in \cite{skenderis} that the ratio of the viscosity to the entropy density is given by the universal holographic value: $$\eta/s~=~1/4\pi~.$$ In other words, the liquid we are considering has the lowest value of the ratio $\eta/s$ possible.

To summarize,  starting from the Euclidean time we were able to elaborate some  field theoretic properties of the medium associated with condensation of a 3d scalar complex field in the Euclidean space. The naive continuation to the Minkowski space fails, however, to preserve the form of the static correlator of  components of the momentum density.  On the other hand, the basic properties of the Minkowskian exotic liquid   \cite{skenderis} do reproduce the field theoretic feature of the liquid we are looking for. The price is that the action (\ref{action}) is highly non-linear while the standard approach (see equations (\ref{son})) is based on the standard version of the  scalar-field action, quadratic in the fields.   Since we treat the time derivatives of the scalar fields as  c-numbers anyhow, there is no reason actually to confine ourselves to a renormalizable version of the effective field theory.   Then we are free to use the non-linear action (\ref{action}).

\subsection{Universality, the stretched membrane and possible relation to phenomenology} 

The liquid living close to the horizon $(r_c\sim r_h)$ describes transport of energy and momentum in the deep infrared region.  Because the geometry close to any region is essentially Rindler, we expect that the exotic liquid (\ref{action}), equivalent to Ricci flat solutions, describes the non-perturbative deep infrared aspects of energy-momentum transport in a wide universality class of physical systems which have a gravity dual. Its universal applicability also follows from the membrane paradigm \cite{membranebook} where we have the Israel junction conditions on the stretched membrane:
\begin{equation}
	T^{AB} = \frac{1}{8\pi}\left[ \kappa_{_+}^{AB} - \kappa_{{_+}C}^{C} \ g^{AB}\right]
\end{equation}
\noindent where $T^{AB}$ is the Browne-York tensor and $\kappa_{_+}^{AB}$ the extrinsic curvature on the exterior.

The stretched membrane is in the Rindler region and hence\footnote{We take a Schwarzschild black hole.}:
\begin{equation}\label{gravitynearhor}
	g = \kappa_{_+}^{00}=4\pi\left(T^{00} + T^a_{\ a}\right)
\end{equation}
\noindent where $g$ is the surface gravity.  If the gravity were Newtonian we would expect $g=4\pi T^{00}$.  Close to the horizon, however, the Newtonian approximation breaks down and we have to use the full relativistic result.  Since $T^a_{~a} = 2p$ and $p=g/8\pi$, (\ref{gravitynearhor}) gives $T_{00} = 0$.   So, whereas a Newtonian  membrane carries a surface energy density to produce the surface gravity, for an Einsteinian membrane, it is the surface pressure that produces the surface gravity \cite{membranebook}.  Therefore on the stretched membrane, the liquid has zero energy density.  More generally, one has \cite{membranebook} on the stretched membrane:
\begin{equation}
	T^{00} = -\frac{1}{8\pi} \ \theta
\end{equation}
\noindent where $\theta$ is the expansion of the null geodesics forming the horizon so that in general the liquid has zero or negative energy density.  $\theta = 0$ corresponds with incompressible flow.

We can now speculate as to how this deep infrared liquid relates to phenomenology.  Some tantalizing hints come from the equation of state of the quark-gluon plasma as studied through lattice simulations.  One of the basic quantities to measure is the conformal anomaly, or the trace of the energy-momentum tensor as a function of temperature:
\begin{equation}
	\theta_{~\mu}^{\mu}~=~\epsilon-3p.
\end{equation}
\noindent In case of Yang-Mills, $\theta_{~\mu}^\mu$ is given by the scale anomaly $\theta_{~\mu}^\mu \sim F^2$.  In Ref. \cite{nakamura3}, the trace anomaly was measured separately for various defects for pure Yang-Mills.  Remarkably enough, it was found that the contribution of the stringy or magnetic component is large and negative.  The component described by the above liquid (\ref{action}) could produce such an effect.


\section{Conclusions} 

In this paper, we suggested that the common construction for a relativistic superfluid is in fact not unique. Once one generalizes non-relativistic superfluidity to the relativistic case the field content of the theory is drastically changed. Namely, one introduces an extra $U(1)$ symmetry and an extra, charged  scalar field. Condensation of the scalar field results in a Goldstone particle. Within the now common approach one matches non-relativistic and relativistic cases by adjusting the chemical potential to be (roughly) equal to the mass. We argue that there exists another generalization, suitable in case of vanishing chemical potential associated with the $U(1)$ charge. The properties of the superfluid component are imitated then by the exotic liquid introduced in Ref. \cite{skenderis}, with vanishing density and dominating pressure.   

We argued that this liquid lives on the stretched membrane and describes energy-momentum transport in the deep infrared.  We also argued that effective $U(1)$ symmetries are inherent to the dual models  for Yang-Mills fields in the Euclidean space-time  and dwelled on phenomenological manifestations of such symmetries and the exotic liquid. \\

\noindent {\bf Acknowledgments}\\
We are thankful to Leo Stodolsky for valuable discussions  and bringing paper \cite{stodolsky} to our attention.  VIZ acknowledges partial support from Federal Special-Purpose Programme 'Cadres' of the Ministry of Education and  Science of the Russian Federation.   

\bibliographystyle{aipproc}   

\begin{thebibliography}{99}

\bibitem{teaney}  E. V. Shuryak,  {\it ``What RHIC experiments and theory tell us about properties of quark-gluon plasma?''}  { Nucl. Phys.}   {\bf 750}  (2005) 64, [arXiv:hep-ph/040506];\\  Th. Schafer, and  D. Teaney, {\it``	 Nearly Perfect Fluidity: From Cold Atomic Gases to Hot Quark Gluon Plasmas ``} { Rept. Prog. Phys. }  {\bf 72}  (2009)  126001,  arXiv:0904.3107 [hep-ph]. 

\bibitem{son1}  P. Kovtun, D.T. Son, A.O. Starinets, {\it ``Viscosity in strongly interacting quantum field theories from black hole physics ``} { Phys. Rev. Lett.} {\bf 94} (2005) 111601, [arXiv:hep-th/0405231];\\D. T. Son, A. O. Starinets, {\it ``Viscosity, Black Holes, and Quantum Field Theory"},  {  Ann. Rev. Nucl. Part. Sci.} {\bf 57} (2007)   95, arXiv:0704.0240 [hep-th]. 

\bibitem{romatschke} P. Romatschke, and U. Romatschke,  {\it ``Viscosity Information from Relativistic Nuclear Collisions: How Perfect is the Fluid Observed at RHIC? ``}  { Phys. Rev. Lett. } {\bf 99}  (2007) 172301, arXiv:0706.1522 [nucl-th]. 

\bibitem{herzog1} S. A. Hartnoll,  {\it ``Lectures on holographic methods for condensed matter physics"}  { Class. Quant. Grav} {\bf 26} (2009) 224002,   arXiv:0903.3246 [hep-th] ;\\Ch. P. Herzog,  {\it ``Lectures on Holographic Superfluidity and Superconductivity"}   { J. Phys.} {\bf A42} (2009) 343001, arXiv:0904.1975 [hep-th].    

 \bibitem{minwala} Ch. P. Herzog, N. Lisker, P. Surowka, A. Yarom, {\it ``Transport in holographic superfluids"}   arXiv:1101.3330 [hep-th];\\  J. Bhattacharya, S. Bhattacharyya, Sh. Minwalla,  {\it Dissipative Superfluid dynamics from gravity},  	arXiv:1101.3332v1 [hep-th];\\  Shu Lin,  {\it ``On the anomalous superfluid hydrodynamics"},  arXiv:1104.5245 [hep-ph].  

  \bibitem{yarom} Ch.P. Herzog, and A. Yarom,  {\it ``	 Sound modes in holographic superfluids ``}  { Phys. Rev.} {\bf D80} (2009) 106002, arXiv:0906.4810 [hep-th]. 

\bibitem{superfluidity}	  D.T. Son, {\it ``Hydrodynamics of relativistic systems with broken continuous symmetries "}, {Int J. Mod. Phys.} {\bf A16S1C} (2001) !284, [arXiv:hep-ph/0011246];\\ M.A. Valle,    {\it ``Hydrodynamic fluctuations in relativistic superfluids ``} { Phys. Rev.} {\bf D77} (2008) 025004, arXiv:0707.2665 [hep-ph];\\ S. A. Hartnoll, P.K. Kovtun, and D.T. Son, {\it``Holographic model of superfluidity"},	 {\it Phys Rev} {\bf D.79} 066002 [arXiv:0809.4870]  

 \bibitem{verschelde}   H. Verschelde, V.I. Zakharov, {\it ``Two-component quark-gluon plasma in stringy models ``}  arXiv:1012.4821 [hep-th];

\bibitem{polchinskibook} J. Polchincki, {\it ``String Theory''} Cambridge Monographs on Mathematical Physics.   

 \bibitem{kogan} B. Sathiapalan,  {\it ``Vortices on the String World Sheet and Constraints on Toral Compactification ``}  { Phys. Rev.} {\bf D35}  (1987)  3277 ; Ya. I. Kogan,  {\it ``Vortices on the World Sheet and String's Critical Dynamics ``}  { JETP Lett.} {\bf 45} (1987)  709. 

\bibitem{atick} J. J. Atick, and E. Witten,  {\it ``The Hagedorn Transition and the Number of Degrees of Freedom of String Theory ``} {  Nucl. Phys.} {\bf B310} 291 (1988).  

\bibitem{polchinski} G. T. Horowitz,   J. Polchinski, {\it ``Selfgravitating fundamental strings"}  {  Phys. Rev.} {\bf D57} (1988) 2557, [arXiv:hep-th/9707170].  

\bibitem{rabinovici} J.L.F. Barbon,   E. Rabinovici,  {\it ``Closed string tachyons and the Hagedorn transition in AdS space"}  {JHEP}  {\bf 0203} (2002) 057, [arXiv:hep-th/0112173]. 

\bibitem{silverstein} A. Adams,   X. Liu, J. McGreevy, A. Saltman, E. Silverstein, {\it ``Things fall apart: Topology change from winding tachyons"}   {JHEP} {\bf 0510} (2005) 033, [arXiv:hep-th/0502021];\\ G. T. Horowitz,  {\it "Tachyon condensation and black holes"} {JHEP}  {\bf 0508} (2005) 091;\\ G. T. Horowitz,   E. Silverstein, {\it ``The Inside story: Quasilocal tachyons and black holes"}   { Phys. Rev.} {\bf D73} (2006) 064016,  [arXiv:hep-th/0601032].  

\bibitem{skenderis} G. Compere,  , P. McFadden,  K. Skenderis,   M. Taylor, {\it ``The holpgraphic fluid dual to vacuum Einstein gravity"} arXiv:1103.3022 [hep-th]  

\bibitem{keeler} I. Bredberg, C. Keeler, V. Lysov, A, Strominger, {\it ``From Navier-Stokes To Einstein"} arXiv:1101.2451 [hep-th]   

\bibitem{lebedev} I.M. Khalatnikov, and V.V. Lebedev, {\it ``Equivalence of two theories of relativistic superfluid mechanics ``}  { Phys. Lett.} {\bf 91A} (1982) 70, { Sov. Phys. JETP} {\bf 56} (1982) 923. 

\bibitem{carter} B. Carter, I. M. Khalatnikov,  {\it ``Canonically covariant formulation of Landau's Newtonian superfluid dynamics ``} {Phys. Rev.} {\bf D45} (1992) 4536, { Ann. Phys.} {\bf 219} (1992) 243.

\bibitem{stodolsky} L.  Stodolsky, {\it ``Decoherence - Fluctuation Relation and Measurement Noise'''}  {  Phys. Rep.} {\bf 320} (1999) 51, arXiv:quant-ph/9903072.   

\bibitem{shirkov} D.V. Shirkov, {\it ``60 years of Broken Symmetries in Quantum Physics: From the Bogoliubov Theory of Superfluidity to the Standard Model"}   { Phys. Usp.} {\bf 52} (2009) 549, { Mod. Phys. Lett.} {\bf A24} (2009) 2802, arXiv:0903.3194 [physics.hist-ph]. 

\bibitem{herzog2} Ch. P. Herzog,  	{\it ``An Analytic Holographic Superconductor"}  {Phys. Rev.} {\bf D81} (2010) 126009,  arXiv:1003.3278 [hep-th].   

\bibitem{kruczenski} M. Kruczenski, and  A. Lawrence,  {\it ``Random walks and the Hagedorn transition ``}  {JHEP}  {\bf 0607} 031 (2006), [arXiv:hep-th/0508148].   

\bibitem{alvarezgaume} L. Alvarez-Gaume,  P. Basu,   M. Marino,  S. R. Wadia, {\it ``Blackhole/String Transition for the Small  Schwarzschild Blackhole of AdS(5)x S**5 and Critical Unitary Matrix Models"}  { Eur. Phys. J.}  {\bf C48} (2006) 647, [arXiv:hep-th/0605041].   

\bibitem{witten} E. Witten, {\it ``Anti-de Sitter space, thermal phase transition, and confinement in gauge theories"}   { Adv. Theor. Math. Phys.} {\bf 2} (1998) 505, [arXiv:hep-th/9803131];\\ E. Witten,  	{\it ``Theta dependence in the large N limit of four-dimensional gauge theories"}  { Phys. Rev. Lett.} {\bf  81} (1998) 2862, [arXiv:hep-th/9807109];\\ N. Itzhaki, J. M. Maldacena, J. Sonnenschein, Sh. Yankielowicz, {\it ``Supergravity and the large N limit of theories with sixteen supercharges"},   {Phys. Rev.} {\bf  D58} (1998) 046004, [arXiv:hep-th/9802042]; \\ T. Sakai, Sh.  Sugimoto,  	{\it ``Low energy hadron physics in holographic QCD"},  {Prog. Theor. Phys.} {\bf 113} (2005) 843, [arXiv:hep-th/0412141];\\ O. Aharony,  J. Sonnenschein, Sh. Yankielowicz, {\it ``A Holographic model of deconfinement and chiral symmetry restoration"}  { Annals Phys.} {\bf 322} (2007) 1420, [arXiv:hep-th/0604161].   

\bibitem{pervye} {\it ``Holographic U(1)(A) and String Creation"} O. Bergman,   G. Lifschytz,  {JHEP} {\bf 043} (2007) 0704. [arXiv:hep-th/0612289]. 

\bibitem{gorsky} A. S. Gorsky, V. I. Zakharov, A. R. Zhitnitsky, {\it ``On Classification of QCD defects via holography ``},  { Phys. Rev.} {\bf D79}  (2009) 106003,  arXiv:0902.1842 [hep-ph]. 

\bibitem{gorsky2} A. Gorsky,   V. Zakharov,  {\it ``Magnetic strings in Lattice QCD as Nonabelian Vortices ``}  { Phys. Rev.} {\bf D77}  (2008) 045017,  arXiv:0707.1284 [hep-th]. 

\bibitem{nakamura} M.N. Chernodub, A. Nakamura, V.I. Zakharov, {\it ``Deconfinement phase transition in mirror of symmetries ``} Proceedings of the Steklov Institute of Mathematics {\bf 272} (2011).    arXiv:0904.0946 [hep-ph].    

\bibitem{pisarski} A. Dumitru, Yun Guo, Y. Hidaka, Ch. P. Korthals Altes, R. D. Pisarski, 	{\it ``How Wide is the Transition to Deconfinement?"} { Phys. Rev.} {\bf  D83} (2011) 034022,  arXiv:1011.3820 [hep-ph]. 

\bibitem{deforcrand} Ph. de Forcrand, A. Kurkela, A. Vuorinen, {\it ``Center-Symmetric Effective Theory for High-Temperature SU(2) Yang-Mills Theory"},  { Phys. Rev. } {\bf D77} (2008) 125014, arXiv:0801.1566 [hep-ph]. 

\bibitem{chernodub2} M.N. Chernodub, V.I. Zakharov, {\it ``Towards understanding structure of the monopole clusters"},  { Nucl. Phys.} {\bf  B669} (2003) 233,  [arXiv:hep-th/0211267]. 

\bibitem{nakamura3} M.N. Chernodub, A. Nakamura and V.I. Zakharov, 	 {\it ``Manifestations of magnetic vortices in equation of state of Yang-Mills plasma"},  Phys. Rev. {\bf D78} (2008) 074021

\bibitem{membranebook} K.S. Thorne, R.H. Price, D.A. Macdonald, {\it ``Black holes and the membrane paradigm"}, Yale University Press, (1986)

\end{thebibliography}










\IfFileExists{\jobname.bbl}{}

 {\typeout{}

  \typeout{******************************************}

  \typeout{** Please run "bibtex \jobname" to optain}

  \typeout{** the bibliography and then re-run LaTeX}

  \typeout{** twice to fix the references!}

  \typeout{******************************************}

  \typeout{}

 }

\end{document}